\documentstyle[psfig]{caosp}
\def\LB{$\lambda$\,Bootis}
\begin{document}
\hauthor{U. Heiter}
\title{Atmospheric parameters and abundances of \LB\ stars}
\author{U. Heiter}
\institute{Institute for Astronomy, University of Vienna, T\"urkenschanzstr. 17,
A-1180 Vienna, Austria}
\date{\today}
\maketitle

\begin{abstract}
The results of the abundance analyses of six \LB\ stars are presented.
For three stars, the impact of individual ODFs and various treatments of
convection in the calculactions of the model atmospheres are investigated.
\end{abstract}

\section{Introduction}
\label{intr}
As has been discussed in the course of this workshop, the classification of
\LB\ stars is not as simple and straightforward as for other groups of
chemically peculiar stars. When only low resolution classification spectra 
are used, it is likely that stars with different astrophysical properties
but similar characteristics in the narrow spectral range are included.
Therefore a detailed spectroscopic investigation for a large number
of confirmed \LB\ stars (e.g. Paunzen et al. 1997) 
covering the whole range of atmospheric parameters 
is required to establish a common abundance pattern for these stars.
Up to now we have analyzed the chemical composition of six \LB\ stars
where two of them turned out to be spectroscopic binaries (Paunzen et al.
1998). 

In order to put our results on more solid ground we investigated the
impact of using different programmes for the calculation of the model
atmospheres on the abundances of three stars with different temperatures,
gravities and metallicities (Heiter et al. 1998). The two issues we examined
concerned the inclusion of individual line opacities and the calculation of the
convective flux in the model atmospheres.

\section{Observations and Analysis}
\label{obse}
\begin{figure}[hbtp]
\psfig{figure=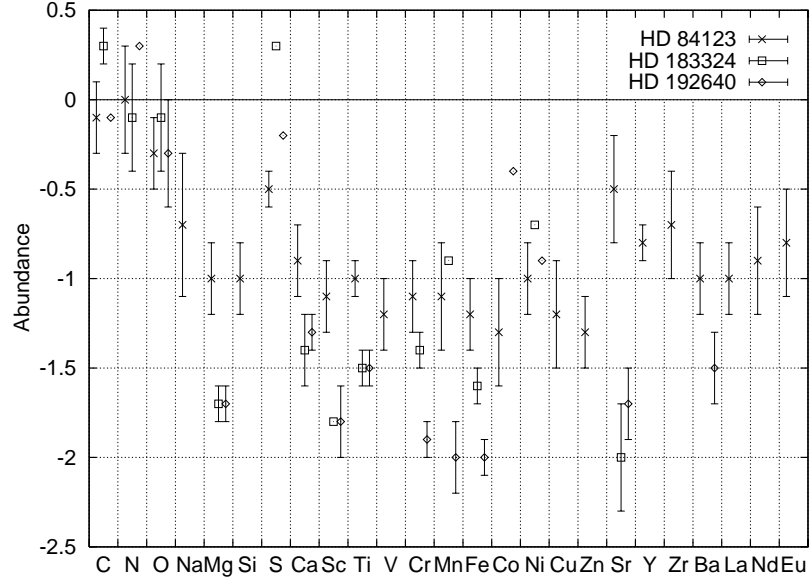,height=8cm}
\caption{Logarithmic abundances relative to the sun for three programme stars.
         Points without error bars indicate upper limits for the abundances.}
\label{abu_fig1}
\end{figure}

\begin{figure}[hbtp]
\psfig{figure=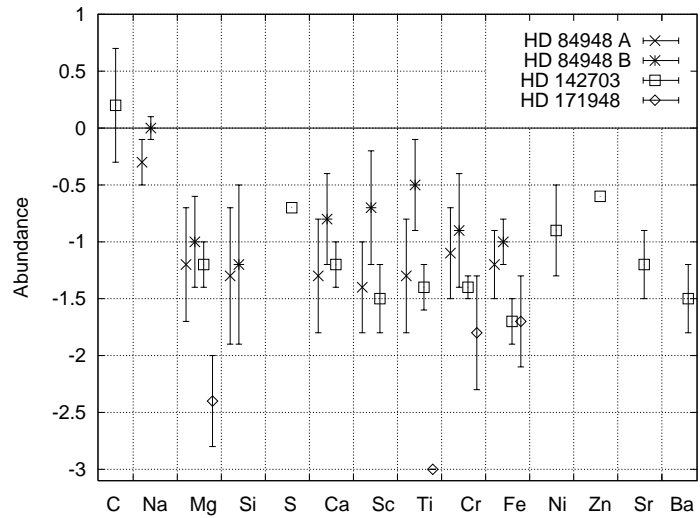,height=7cm}
\caption{Same as Fig.~\ref{abu_fig1} for the remaining programme stars.}
\label{abu_fig2}
\end{figure}

High resolution spectra with high signal-to-noise ratios have been obtained at 
various sites as listed in Table~\ref{obs_tab}. They have been reduced with NOAO
IRAF (Willmarth \& Barnes 1994), and atmospheric parameters and abundances
have been derived as described by Gelbmann et al. (1997, 1998). The resulting
atmospheric parameters are presented in Table~\ref{par_tab}. The abundances
of each element for which at least one ``unblended'' line could be detected
in the observed spectra of the programme stars are plotted in 
Figs.~\ref{abu_fig1} and \ref{abu_fig2}. ``Unblended'' means that no other line 
with a depth greater than 30~\% of that of the examined line is found 
within a wavelength window whose width depends mainly on the $v \sin i$ of 
the respective star.

\begin{table}[t]
\small
\caption{Observations used for this analysis.}
\label{obs_tab}
\begin{center}
\begin{tabular}{|l|l|l|l|l|l|}
\hline
{\bf HD}&{\bf Observatory}&{\bf [m]}&{\bf Spectrograph}&{\bf Date}
&{\bf Observer}\\
\hline
{\bf 84123} & Asiago (Italy) & 1.8 & Echelle & 3--1995, & Heiter \\ 
{\bf 84948} & & & & 2--1997 & \\
\hline
{\bf 142703} & OPD/LNA (Brazil) & 1.6 & Coud\'e & 6--1995 & Paunzen \\
\hline
{\bf 171948} & McDonald (Texas) & 2.1 & Sandiford & 4--1996 & Handler \\ 
 & & & Echelle & & \\
\hline
{\bf 183324} & OHP (France) & 1.5 & Aurelie & 6/8--1994 & Gelbmann \\
{\bf 192640} & & & & & +Kuschnig \\
\hline
\end{tabular}

\vspace{1cm}

\begin{tabular}{|l|l|l|l|}
\hline
{\bf HD} & {\bf Wavelength range [\AA]} & {\bf Resolution} & {\bf S/N} \\
\hline
{\bf 84123} & 4000$-$7200 & 30000 & 150 \\ 
{\bf 84948} & & & \\
\hline
{\bf 142703} & 4100$-$4900 & 30000 & 100 \\
\hline
{\bf 171948} & 4290$-$4730 & 60000 & 150 \\ 
\hline
{\bf 183324} & 3800$-$5300, 5800$-$6000 & 20000 & 200 \\
{\bf 192640} & +5900$-$6280, 7050$-$7230 & & \\
\hline
\end{tabular}
\end{center}
\end{table}

\begin{table}[t]
\small
\caption{Atmospheric parameters of the programme stars.}
\label{par_tab}
\begin{center}
\begin{tabular}{lccccc}
\hline
\hline
{\bf star}&{\bf T$_{\rm eff}$}&{\bf log g}&{\bf v$_{\rm turb}$}
&{\bf [Z]}&{\bf v sin i}\\
{\bf HD}&{\bf ($\pm$200 K)}&{\bf ($\pm$0.3)}&{\bf ($\pm$0.5\,km\,s$^{-1}$)}& 
 & {\bf [km\,s$^{-1}$]} \\
\hline
{\bf \ 84123} & 6800  & 3.5 & 3.0  & $-$1.0 & 15(5) \\ 
{\bf \ 84948A} & 6600 & 3.3 & 3.5  & $-$1.0 & 45(5) \\
{\bf \ 84948B} & 6800 & 3.7 & 3.5  & $-$1.0 & 55(5) \\
{\bf 142703} & 7000 & 3.7 & 3.0  & $-$1.5 & 100(10) \\
{\bf 171948A} & 9000 & 4.0 & 2.0  & $-$2.0 & 15(5)  \\
{\bf 171948B} & 9000 & 4.0 & 2.0  & $-$2.0 & 10(5)  \\
{\bf 183324} & 9300  & 4.3 & 3.0  & $-$1.5 & 90(10) \\
{\bf 192640} & 7800  & 4.0 & 3.0  & $-$2.0 & 80(10) \\
\hline
\hline
\end{tabular}
\end{center}
\end{table}


\section{Self-consistent model atmospheres}
\label{self}
The abundances of the three stars from Fig.~\ref{abu_fig1} were used to
calculate individual opacity distribution functions (ODFs) for these stars with
a programme written by N.~Piskunov (Piskunov \& Kupka 1998). The subsequent
inclusion of these ODFs in the model atmosphere calculations instead of
the tabulated opacities from Kurucz (1993) showed that the $T(\tau)$-relations
and therefore the synthetic spectra of all three stars do not change
significantly.

\vspace{-3mm}
\begin{figure}[hbtp]
\psfig{figure=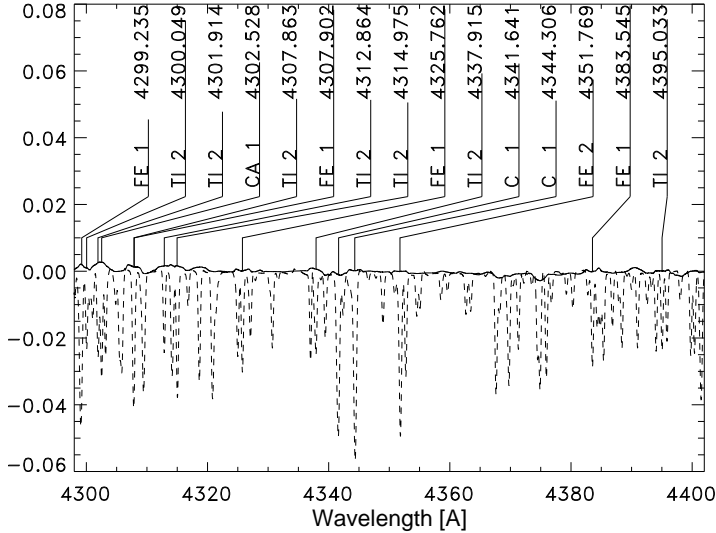,height=8cm}
\caption{Difference between normalized synthetic spectra calculated with CM
and MLT. The solid and dashed lines represent HD\,192640 and HD\,84123,
respectively.}
\label{con_fig1}
\end{figure}
\begin{figure}[hbtp]
\psfig{figure=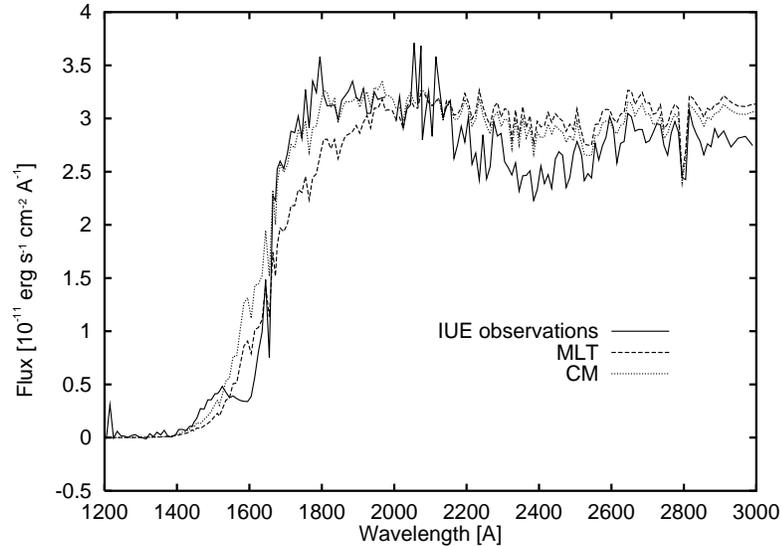,height=7.5cm}
\caption{Comparison of synthetic surface fluxes with IUE observations
for HD\,192640.}
\label{con_fig2}
\end{figure}

\vspace{-3mm}
In the next step we compared the results of the original ATLAS9
programme (Kurucz 1993, MLT) with three versions where the treatment of
convection (or overshooting) had been changed: 1) Deletion of the part of
the programme concerning the overshooting (NO), 2) Implementation of the
convection model by Canuto \& Mazzitelli (1991, CM), 3) Variation of the
treatment of overshooting (Castelli 1996, OVOK). The differences can
be summarized as follows: Using the NO version results in similar changes
compared to MLT as for the CM model, whereas OVOK yields practically the same
$T(\tau)$-relation as MLT. For HD\,84123, the largest differences occur in
the visible spectral range (see Fig.~\ref{con_fig1}), where the abundances 
of all elements would have to be decreased by 0.1~dex to match the MLT 
calculations. For HD\,192640, significant differences are only seen in the
UV spectral range (Fig.~\ref{con_fig2}). The CM flux seems to represent the
observations better than that of MLT. Note the absorption feature at 1600\,\AA,
which cannot be reproduced by either of the models.
For the hottest of the stars, HD\,183324, the differences are negligible.

\acknowledgements
This investigation was carried out within the working group {\em
Asteroseismology--AMS} with funding from the Fonds zur F\"orderung
der wissenschaft-lichen Forschung (project {\em S7303-AST}). 
We thank Dr.~E.~Solano for providing the IUE spectra.
Use was made of the Simbad database, operated at CDS, Strasbourg, France.


\begin{thebibliography}{}
\article{Canuto, V.M., Mazzitelli, I.}{1991}{\apj}{370}{295}
\bibitem{} Castelli, F.: 1996, in {\it Model Atmospheres and Spectrum 
Synthesis},
 eds.: S.J.~Adelman, F.~Kupka and W.W.~Weiss, ASP Conf. Ser. 108, p. 85
\bibitem{} Gelbmann, M.: 1998, {\it this workshop}
\article{Gelbmann, M., Kupka, F., Weiss, W.W., Mathys, G.}{1997}{\aaa}{319}{630}
\article{Heiter, U., Kupka, F., Paunzen, E., Weiss, W.W., Gelbmann, M.}{1998}
{\aaa}{}{submitted}
\article{Paunzen, E., Heiter, U., Handler, G., Garrido, R., Solano, E.,Weiss,
W.W., Gelbmann, M.}{1998}{\aaa}{329}{155}
\article{Paunzen, E., Weiss, W.W., Heiter, U., North, P.}{1997}{\aaas}{123}{93}
\article{Piskunov, N., Kupka, F.}{1998}{\aaa}{}{in preparation}
\bibitem{} Kurucz, R.L.: 1993, {\it CD-ROM 1-23}, Smithsonian Astrophysical
Observatory
\bibitem{} Willmarth, D., Barnes, J.: 1994, {\it A User's Guide to Reducing
Echelle Spectra with IRAF}, NOAO
\end{thebibliography}
\end{document}